\documentclass[twocolumn,showpacs,amsmath,amssymb]{revtex4}

\usepackage{graphicx}
\usepackage{dcolumn}
\usepackage{bm,color}

\begin{document}

\preprint{APS/}

\title
[Conditional recovery of time-reversal symmetry in many nucleus systems]
{Conditional recovery of time-reversal symmetry in many nucleus systems}

\author{Yoritaka Iwata$^{1}$}
 \email{iwata_phys@08.alumni.u-tokyo.ac.jp}
\author{Paul Stevenson$^{2}$}

 \affiliation{$^{1}$Institute of Innovative Research, Tokyo Institute of Technology, Japan}
 \affiliation{$^{2}$Department of Physics, University of Surrey, United Kingdom}
\date{\today}

\begin{abstract}
The propagation of non-topological solitons in many-nucleus systems is studied based on time-dependent density functional calculations, focusing on mass and energy dependence.
The dispersive property and the nonlinearity of the system, which are inherently included in the nuclear density functional, are essential factors to form a non-topological soliton.
On the other hand, soliton propagation is prevented by charge equilibration, and competition can appear between soliton formation and disruption.
In this article, based on the energy-dependence of the two competitive factors, the concept of conditional recovery of time-reversal symmetry is proposed in many-nucleus systems.
It clarifies the possibility of preserving the nuclear medium inside natural or artificial nuclear reactors, at a suitable temperature.
From an astrophysical point of view, the existence of the low-temperature solitonic core of compact stars is suggested.
\end{abstract}


\maketitle

\section{Introduction}
Dispersion and nonlinearity are essential to the existence of non-topological solitons, where a non-topological soliton means a wave propagating with perfect transparency.
The nonlinear Schr\"{o}dinger equation (for a textbook, see \cite{11ablowitz})
\[
 i \partial_t \psi(x,t) =  - \partial_x^2 \psi(x,t) - 2 g | \psi(x,t)|^2 \psi(x,t)
\]
is a typical example exhibiting bright solitons for $g>0$ in which the first and the second terms of the right hand side are responsible for the dispersive property and nonlinearity, respectively.

From a physical point of view, soliton propagation in quantum systems depends substantially on the collective degree of freedom.  In the case of colliding atomic nuclei, fast charge equilibration \cite{10iwata}, which is a kind of quantum and fermionic many-body effect, prevents soliton propagation.
Indeed the formation of a uniform charge distribution is led by a charge equilibrating wave, while the preservation of a non-uniform density distribution is conferred by the soliton wave.  In this example, it is the wave function of each of the colliding nuclei that may be considered a soliton, if the nuclei pass through each other and remain unperturbed by the collision.

According to a comparison to other nuclear theories and experiments \cite{18zhu,17zhang,17zhu,14souliotis,17umar}, an almost universal propagation speed several tens of percent slower than the fermi velocity (corresponding to almost 25\% of the speed of light for many nucleon systems) is found for the fast charge equilibration, and it is suggested to be associated with the zero sound propagation \cite{75leggett} of the many-nucleon system \cite{12iwata}.
This propagation speed is higher than the relative velocities of low-energy ion collisions.
Since the soliton propagation speed is given by the relative velocity of collisions of the two ions, the soliton wave propagation is expected to be effectively prevented by the fast charge equilibrating wave in low-energy ion collisions.
This is an intuitive explanation for the mechanism for the energy-dependent existence of soliton propagation \cite{15iwata} in nuclei, where there is an upper energy limit for the appearance of the fast charge equilibrating wave \cite{10iwata}.

Non-topological soliton propagation in many nucleon systems should be different in its typical order of propagation speed compared with other cases such as soliton propagation in many quark systems.
In addition the nuclear symmetry energy is expected to play a role in soliton propagation, as it is the main driving force of the fast charge equilibration~\cite{17stone}.

In this article a realization of the mathematical concept of non-topological solitons is presented in sub-atomic fermionic systems. Some special conditions must be satisfied for the existence of the soliton in a system of interacting nuclei, because the medium consists of two components: protons and neutrons.
In this sense non-topological soliton propagation in many-nucleus systems is found to be essentially different from that in optics (massless single component system) or electron systems (single component system).   
As a result, the conservation inherent to non-topological solitons leads to the conditional recovery of time-reversal symmetry in many-nucleus systems.
Possible applications of non-topological soliton are presented from astrophysics to nuclear engineering.

\section{Nuclear Systems}

The nuclear medium confined in reactors, stars and so on consists of many kinds of nucleus in which the statistical nuclear kinetic energy determines the temperature of the nuclear medium.
Meanwhile, the nucleus is a many-nucleon system, which consists of protons and neutrons, and is a nonlinear system interacting by the nuclear and the Coulomb forces. 
In order to understand soliton propagation in many-nucleus systems, they are ideally studied by including detailed nucleon degrees of freedom.
For the existence of non-topological soliton propagation, the competition between the nonlinear transparency and the charge equilibration is essential at the level of nucleon degrees of freedom. 
Those two mechanisms compete with each other; indeed the former one prevents mixing the two species of nucleon while the latter one contributes to form a mixed nuclear medium with a certain neutron-to-proton number ratio ($N/Z$ ratio).
Temperature control is expected to be possible if the conditions from the two mechanisms allow a meeting energy range.

\begin{figure} 
\includegraphics[width=80mm]{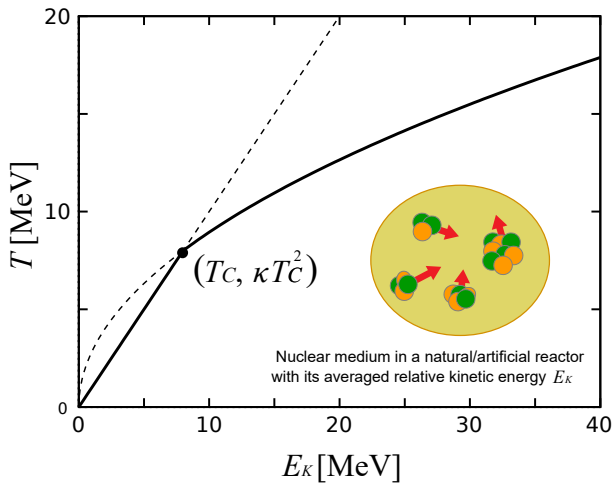} 
\caption{ \label{fig5}
(Color online) A model for many-nucleus system is shown by a thick line.
Temperature is estimated by Eq.~(\ref{temp}) with $T_C = 7.2$~MeV.
$E_K$ is the relative kinetic energy of a representative nucleus in the many-nucleus system.
Note that $(T_C, \kappa T_C^2)$ always satisfies Eq.~(\ref{temp}) independent of the choice of $T_C$ and $\kappa$.
In the present examination soliton propagation is studied for low-energy medium with temperature below 30~MeV.
The inserted figure illustrates nuclear medium in our scope, where each nucleus is made of protons and neutrons, and the momentum vectors of nuclei are shown by red arrows.
}
\end{figure}

An order estimation for the temperature of many-nucleus systems (not of many-\textit{nucleon} systems) is necessary, and it is provided by following the manner used in many-nucleon system.
Employing the definition of the nuclear temperature using the Bethe formula \cite{37bethe, 96froebrich}, the temperature of a many-nucleus system is estimated in terms of the relative kinetic energy $E_K$ by
\begin{equation} \label{temp} \begin{array}{ll}
E_K = 
\left\{ \begin{array}{ll}
  \kappa T_C  T, \quad T \le T_C  \vspace{2.5mm} \\
  \kappa T^2, \quad T > T_C
 \end{array} \right.
\end{array} \end{equation}
where two parameters are included.  The critical temperature $T_C$ is introduced for a practical reason to distinguish the different energy dependence of temperature between high and low energies, and it is reasonable to take $T_C = \kappa^{-1} = 7.2$~MeV \cite{10iwata}.
Note that $T_C = 7.2$~MeV is associated with the translation of the fermi energy of many-nucleon system to the relativistic center-of-mass kinetic energy \cite{13iwata-02}. 
In low energies, a real constant $\kappa$ corresponds to a proportional constant between $E_K$ and $T_C$ (cf. the Boltzmann factor).
$E_K$ behaves linearly for low temperature and quadratically for high temperature (Fig.~\ref{fig5}).
In order to define the temperature, it is necessary to know the relative kinetic energy $E_K$ between nuclei.
Here we assume the statistical relative kinetic energy is well approximated by the collision energy of given representative heavy-ion collisions.
Although the temperature can be defined in a thermalized statistical ensemble, it is not achieved at each moment of heavy-ion collisions. Here, by assuming that the temperature is unchanged during the heavy-ion collisions, the temperature is defined by the representative relative kinetic energies of initial states.
Such a treatment is sufficient to have an order estimation for the temperature of nuclear medium.

\begin{table}[t] 
\caption{Self-binding energies of the ground states are calculated using the SV-bas effective interaction.
Binding energy per nucleon [MeV] of the initial states are compared to those of intermediate states, and the corresponding experimental values are shown in parenthesis \cite{nudat2}.
The system for $(A,Z)=(204, 100)$ is too proton-rich to have a bound state.
}
\begin{center}
  \begin{tabular}{l|c|c|c}  \hline \hline
 (A,Z)  &  $B(^{A}Z)$ &  $B(^{A+4}Z)$ &  $B(^{2A+4}2Z)$         \\ \hline 
 (4,2) &  $6.93$ (7.08)  & $4.48~(3.93)$ & $6.07$~(7.06) \\
 (16,8)  & ~$8.21$ (7.98)~   & ~$7.84$ (7.57)~  & ~$8.75$ (8.58)~  \\
 (40,20) & $8.73$ (8.55)  & $8.80$ (8.66) & $8.59$ (8.36) \\
 (48,20) & $8.83$ (8.67)  & $8.59$ (8.43) & $ 8.60$ (8.52) \\
 (100,50) & $8.35$ (8.25)  & $8.45$ (8.38) & ~---~ (~---~) \\
 (120,50) & $8.54$ (8.50)  & $8.51$ (8.47) & $7.47$ (7.47) \\
 \hline \hline
\end{tabular} \label{table-binding}
\end{center}
\end{table}

\section{Heavy-Ion Reactions}

Representative ion reactions
\begin{equation} \label{reaction-gen}
 ^{A+4}Z ~+~ ^{A}Z ~\to~  \vspace{1.5mm} 
\end{equation}
in the nuclear medium are considered.
In fact reactions of six kinds $(A,Z)$ = (4,2), (16, 8), (40, 20), (48, 20), (100, 50) and (120, 50) are investigated for clarifying the conservation properties of the nuclear medium.
Different $N/Z$ pattern formation of the nuclear medium is expected to appear depending on the combination of $Z$ and $N$, and the soliton propagation, if any, should be altered accordingly.
The choice of these combinations follows from a preceding research on nuclear symmetry energy~\cite{17stone}  in which the highest achievable density and the corresponding asymmetry parameter $\delta$ have been presented for $(A,Z)$ = (40, 20), (48, 20), (100, 50) and (120, 50)  cases.
This examination covers a mass range from 12 to 244.
It is notable that reactions of this kind recently received special attention in terms of identifying the extraordinary neutron-rich resonance state of the tetra-neutron~\cite{16kisamori,03pieper,02marques,16hiyama}.
In the present setting, $^{A+4}Z$ of several kinds play a role of mother nuclei for the tetra-neutron.

Central collisions are calculated, since effects of non-central collisions have already been assessed to enhance the soliton property as much as several 10\% \cite{15iwata}.
That is, central collisions bring about the most strict condition for the realization of the soliton property, and this property is then amenable to the present examination of showing a possible application of the soliton propagation in many-nucleus systems.
As a result of these reactions, depending on the collision energy from 1 MeV to several 10s of MeV per nucleon, fusion, deep-inelastic collision, and collision-fission such as fusion-fission and quasi-fission are able to appear.

Table~\ref{table-binding} compares the binding energies of the colliding nuclei to those of intermediate fused systems.
For $Z \ge 20$, the $B(^{2A+4}2Z)$ value becomes smaller compared to $B(^{A}Z)$ and $B(^{A+4}Z)$ values. That is, the change of reaction type from exothermic to endothermic is included in the present range of calculations.
Collision energies with 1 to 100 MeV per nucleons (in the center-of-mass frame) are taken into account.
The collision energy corresponds to the relative kinetic energy, and it is simply called the energy of the system in the following.

As in the preceding research~\cite{15iwata}, the high energy cases around $E_K$ = 100.0~MeV are presented for theoretical interest, because the high energy cases with those relative velocity almost equal to 80\% of the speed of light lie beyond the application limit of non-relativistic theory employed in the following.

\begin{figure}[t] 
\includegraphics[width=85mm]{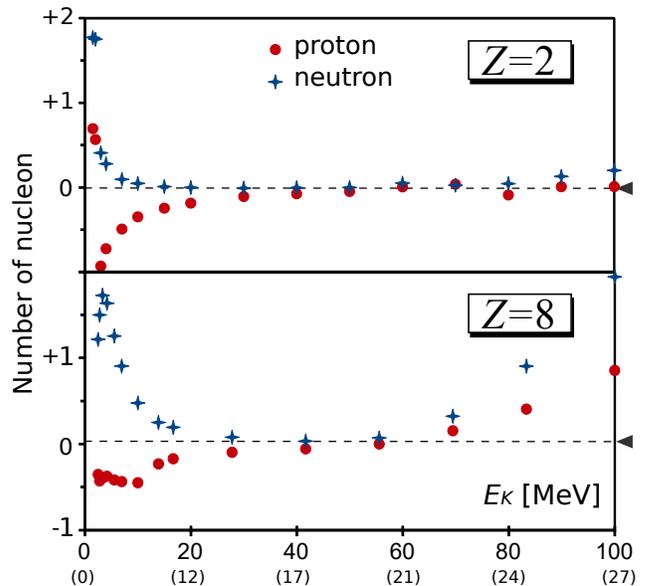} 
\caption{ \label{fig4}
(Color online) Nucleon transfer took place in reactions (\ref{reaction-gen}).
Numbers of transferred protons and neutrons from $^{A+4}Z$ to $^{A}Z$, where $(A,Z)$ = (4, 2) and (16, 8).
The horizontal axis shows the energy per nucleon (temperature [MeV] in the parenthesis), and the perpendicular axis shows the number of transferred nucleon.
The value becomes positive if nucleon is transferred from $^{A+4}Z$ to $^{A}Z$, and negative if nucleon is transferred from $^{A}Z$ to $^{A+4}Z$.
The fusion events are ignored, so that the lowest energies in graphs correspond to a few MeV.
Since the statistical distribution represented by probability wave is calculated in the TDDFT, the calculated nucleon number interpreted as the number probability generally becomes non-integer.}
\end{figure}

\section{Formalism}

Theoretical investigations are carried out based on the time-dependent nuclear density functional theory (TDDFT, for short).
For $r \in \Omega \subset {\mathbb R}^3$ and $t \in {\mathbb R}$, the TDDFT equation is written by
\begin{equation}  \label{mastereq}
  i \hbar \partial_t \psi_{q,j}(r,t) = H (\psi_{q,j}(r,t))  \psi_{q,j}(r,t),
\end{equation}
with 
\begin{equation}
  \begin{aligned}
     H(\psi_{q,j}(r,t))& =
    U_q(r)-\nabla\cdot\left[B_q(r)\nabla\right]
    + i \vec W_q\cdot(\vec\sigma\times\nabla) \\
    &+\vec S_q\cdot\vec\sigma
    -\frac{\mathit{i}}{2} \left[(\nabla\cdot\vec A_q)+2\vec
      A_q\cdot\nabla\right]\,,
  \label{eq:spham}
\end{aligned}
\end{equation}
where $ \psi_{q,j}(r,t)$ denotes single-nucleon wave function ($j = 1,2, \cdots 2A+4$), the index $q$
identifies isospins.  $H (\psi_{q,j}(r,t))$ is the effective Hamiltonian operator derived from nuclear density functional theory.  In Eq.~(\ref{eq:spham}), $\nabla$ is the spatial derivative operator, $\vec\sigma$ is a vector of Pauli spin matrices, and all other symbols are functions derived from the density functional parameters \cite{14maruhn}.  In essence, the Hamiltonian consists of a kinetic energy term, a mean-field potential, a spin-orbit term and terms including time-odd densities which ensure Galilean invariance.  See \cite{72vautherin} for the detail representation, refer to \cite{13iwata} for a mathematically rigorous derivation, and refer to \cite{19stevenson} for a review of recent research on nuclear energy density functional within heavy ion collisions.

Equation \eqref{mastereq} is a nonlinear Schr\"{o}dinger type equation satisfying the dispersive property.  TDDFT is a unique theoretical framework for investigating the solitonic transportation in many-nucleus systems as built upon the nucleon degrees of freedom (for a review of TDDFT including recent developments, see \cite{18simenel}).
Due to the Schr\"{o}dinger type formalism, the TDDFT is the non-relativistic theory whose theoretically-permissible upper limit of the relative kinetic energy is several 10 MeV per nucleon.

Periodic boundary conditions are imposed on both stationary and nonstationary problems.
In particular the stationary problem is numerically solved to prepare the initial state consisting of two colliding nuclei with a certain relative velocity.

As an effective interaction, the SV-bas nuclear interaction \cite{09klupfel} and the Coulomb interaction are implemented.
The SV-bas interaction is especially known for well reproducing the neutron skin thickness of heavy nuclei such as $^{208}$Pb (for a compilation of experimental and theoretical results, see~\cite{15neumann}). 
This property is expected to be relevant for the stability of the tetra-neutron.
Although the pairing interaction is not introduced in the present density functional, the collision energy is sufficient high for pairing interaction not to play a significant role.
Indeed, from an energetic point of view, the nuclear pairing is the effect less than a few 100s of keV per nucleon.
Numerical solutions are obtained based on the finite difference method (for details, see Ref.~\cite{14maruhn}). 
The numerical settings such as time an space unit sizes follow from the preceding research~\cite{15iwata}.
By comparing to experimental nuclear binding energies (Table~\ref{table-binding}), the validity of the present theoretical framework can be recognized, where the difference between theoretical and experimental values are less than 4\% at the most.
The theoretical framework with the numerical settings is concluded to be sufficient to simulate the ion collisions.

\begin{figure*} 
\hspace{-45mm} (a) $Z=20$ cases \hspace{58mm} (b) $Z=50$ cases\\
\includegraphics[width=80mm]{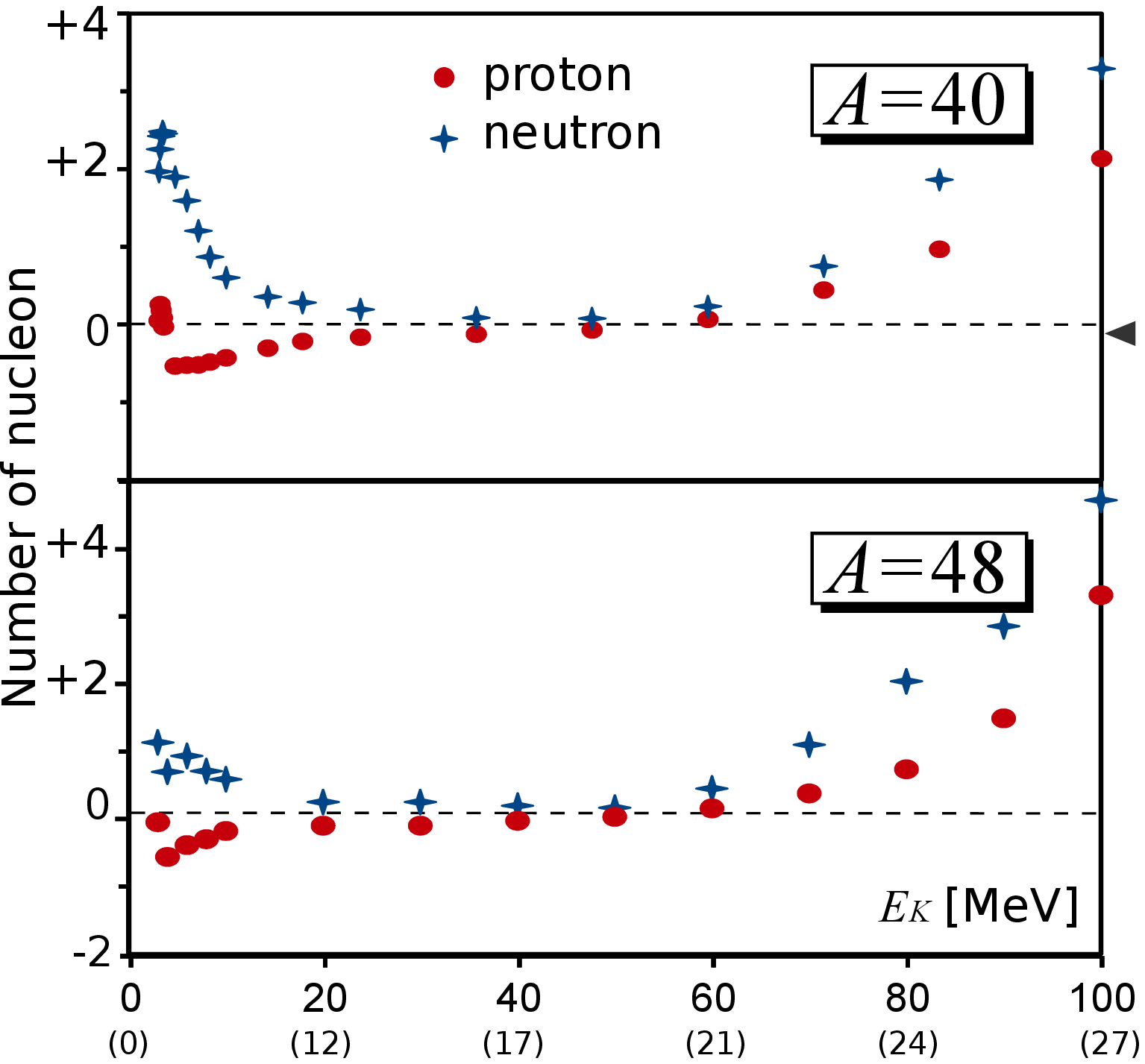} 
\includegraphics[width=80mm]{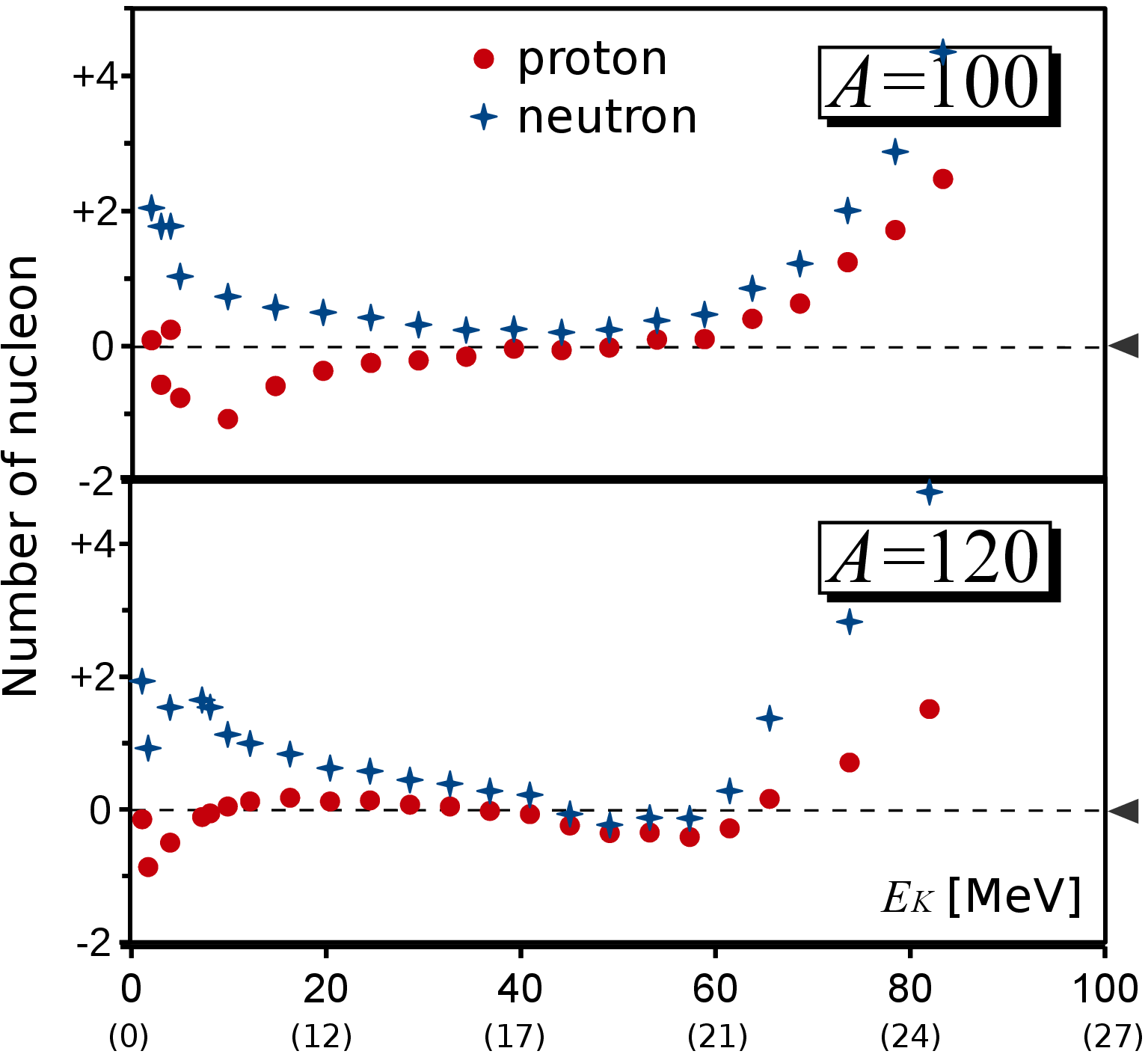} 
\caption{ \label{fig8}
(Color online) Numbers of transferred protons and neutrons from $^{A+4}Z$ to $^{A}Z$, where $Z=20$ is examined in the left, and $Z = 50$ in the right.
The drawing manner follows from Fig.~\ref{fig4}.}
\end{figure*}

\section{Summary of Previous Results}
Before moving on to the main discussion, we briefly review the preceding results~\cite{15iwata}.
According to the preceding research dealing with $^{4}$He+$^{8}$He reactions, the dominance of nonlinear $t_1$-term $|\psi|^2 \psi$ in the density functional has been confirmed.  
Indeed there cannot be any bound states by switching off the $t_1$-term.  Note that the $t_1$-term from the density functional manifests itself across all terms of the terms in (\ref{eq:spham}) except the spin-orbit terms \cite{14maruhn}.
On the other hand, $t_1$-term contribution is highly modified by the secondary-dominant nonlinear fractional power $t_3$-term $|\psi|^{2+\alpha} \psi$ (for the dominance, compare the results with and without $t_1$ and $t_3$ terms in Table 2 of Ref.~\cite{15iwata}).  The $t_3$-term contributes only to the potential $U_q$ in (\ref{eq:spham}).
Dominance of those terms in the nuclear density functional implies the validity of an energy-dependent soliton existence in which $t_1$-term and $t_3$-term are responsible for the soliton existence and energy dependence respectively. 
It is remarkable here that $t_3$-term is known to be indispensable for the reproducing of the incompressibility of nuclear matter and the related excitation states.
In Ref.~\cite{15iwata},  for ion reactions including light ions, the energy-dependence of soliton emergence has been clarified in terms of both mass and momentum transparencies.
A rough sketch of the energy-dependence is as follows: the soliton property is not so active for low energies less than a few MeV per nucleon, soliton property becomes active around 10 MeV per nucleon, it achieves almost the perfect transparency around 20 to 40 MeV per nucleon, and the transparency again decreases for much higher energies.

\section{Calculations and Discussion}

The soliton property is defined by perfect transparency, so that perfect fluidity without dissipation is satisfied by the solitons.
The existence of solitons can be confirmed by the absence of both neutron and proton transfers, and by the absence of internal nucleon excitation during a collision between two nuclei.
If these conditions are satisfied, the shape of colliding nuclei is automatically conserved before and after the collision.
Figure \ref{fig4} shows the nucleon transfer for  $^{8}$He+$^{4}$He and $^{20}$O+$^{16}$O collisions. 
First, the similarity between the two cases are noticed.
Positive neutron transfer and negative proton transfer appear simultaneously at low temperatures less than 10 MeV.
This is classified as a kind of charge equilibration \cite{10iwata-npa}, in which proton-rich $^{A}Z$ gains neutrons and loses protons simultaneously.
Solitonic transportation events are seen at intermediate temperatures from 10 to 20 MeV.
It means that the tetra-neutron independence (the four extra neutrons in the $^{A+4}Z$ system retain their character) is true within a limited energy range.

As prominently seen in case of $Z=8$, massive neutron-proton transfers appear for high temperatures above 21 MeV.
Indeed $^{A}Z$ gains protons with almost the twice numbers of neutrons (transfer with $N/Z \sim 2$).
Not only the charge difference but also the mass difference of the two nuclei becomes smaller in this kind of neutron-proton massive transfer reactions.
Since significant numbers of nucleons are emitted from the final products during or promptly after the collision (cf. multi fragmentation) in high temperatures larger than 20 MeV, these reactions undergo momentum equilibration rather than the mass equilibration.
More precisely, the transportation at higher energies is understood as the mixture of neutron and deuteron emissions carrying away large amounts of momentum.
This is to say, the main stabilization or cooling mechanism  of higher-energy ion collisions is confirmed to be nucleon emissions.
Second, for the mass dependence, the soliton property is more suppressed in the heavier case.
This tendency is not seen in the low temperature side, but prominently in the high temperature side.
The prevention of soliton propagation at lower temperature shows the universal validity of fast charge equilibration mechanism, while the considerable prevention at higher temperature is due to the momentum equilibration mechanism.

The left panels of Fig. \ref{fig8} compare the nucleon transfer between $^{44}$Ca+$^{40}$Ca and $^{52}$Ca+$^{48}$Ca, and the right panels compare the nucleon transfer between $^{104}$Sn+$^{100}$Sn and $^{124}$Sn+$^{120}$Sn. 
Solitonic transportation events are seen at intermediate temperatures about 20 MeV.
For $^{44}$Ca+$^{40}$Ca and $^{52}$Ca+$^{48}$Ca reactions, there is no significant difference except for the amount of massive neutron-proton transfers at higher temperatures.
Note that momentum equilibration is affected by stable bound states; 
$^{40}$Ca is the most stable among $Z=20$ nuclei, and
$^{116}$Sn is the most stable among $Z=50$ nuclei.
On the other hand, in case of $(A,Z)=$ (48, 20), (100, 50), and (120, 50) reactions, $^{50}$Ca,  $^{102}$Sn, $^{122}$Sn are the charge equilibrated nuclei, respectively.
Despite the fact that there is a remarkable difference between the most stable and the charge equilibrated nuclei for a given $Z$, charge equilibration always appears in lower energy side.
Among several features, neutron richness plays a quite limited role in altering the confirmed energy-dependent prevention of soliton propagation.
We have confirmed a universal trend that the soliton property is more suppressed in the heavier case. 

Soliton independence is broken by a nuclear symmetry energy effect. 
As seen in Figs. \ref{fig4} and \ref{fig8}, the soliton breaking mechanism in lower and higher energy sides are different.

It is important to note that the soliton energies are located in between the lower and higher non-soliton energies, so that the existence of a soliton cannot be recognized as simple passing-through events due  to the lack of interactions.  The interactions must be taking place, since they take place either side in energy, but be active in maintaining solitonic behaviour.

Since nuclear symmetry energy is dependent on both its density and temperature, the transportation in extremely high density situations can  be different. 
Although quite a few things are known for nuclear medium at extreme situations at this point, achievable high densities and the corresponding nuclear asymmetries in some heavy-ion collisions can be found in Ref.~\cite{17stone}.
The curvature of $E_K$-dependence in Fig. \ref{fig8} are far from flat for $N/Z= 7/5$ cases ($A=48$ and $A=120$).  Consequently the solitonic transportation can be influenced primarily by the total mass of the colliding system (the size of the colliding nuclei: $A=N+Z$), and secondarily by the nuclear symmetry energy effect (the detail composition of colliding nuclei: $N/Z$). 

\begin{figure}
\includegraphics[width=80mm]{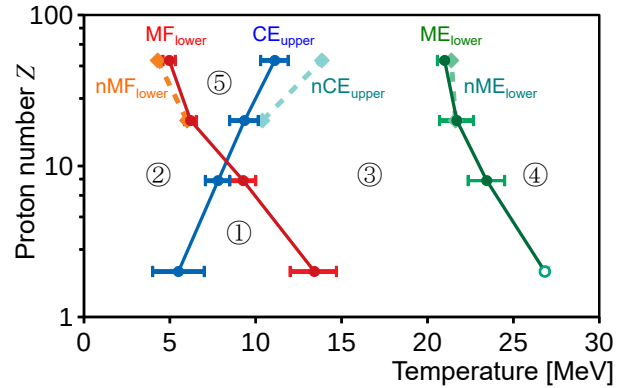} 
\caption{ \label{fig11}
(Color online) Classification of reaction types depending on the temperature and mass.
Charge equilibration upper-limit energy (CE$_{\rm upper}$) is compared to the multi-fragmentation lower-limit energy (MF$_{\rm lower}$) and massive momentum transfer lower-limit energy (ME$_{\rm lower}$).
For $N/Z$=1 cases, the error bars are added by the calculated collision energies, the central points of lower side and upper side of error bars are connected by thick lines, where ME$_{\rm lower}$ with $Z=2$ is calculated to be larger than T= 29 MeV (corresponding to the upper bound of our calculation $E_K = 100$ MeV).
For $Z=20$ and 50 cases, $N/Z$=7/5 cases are plotted with dashed lines and without error bars (for the better comparison without any congestion) that are shown by nCE$_{\rm upper}$, nMF$_{\rm lower}$ and nME$_{\rm lower}$ respectively.
As a result, focusing only on $N/Z$=1 cases, there are five areas labeled from $ \textcircled{\scriptsize 1}$ to $\textcircled{\scriptsize 5}$ separated by the thick lines.
}
\end{figure}

Figure~\ref{fig11} summarizes all the calculated reactions.
The boundary energy between charge equilibration calculated in the lower energy and the solitonic transportation is shown by CE$_{\rm upper}$.
The boundary energy between large momentum transfer calculated in the higher energy and the solitonic transportation is shown by ME$_{\rm lower}$.
The boundary energy indicating whether nucleon emission or multi-fragmentation of more than two fragments appears or not is shown by MF$_{\rm lower}$.
In all the boundaries, the neutron or proton is identified to be transferred if the calculated transferred nucleon number is more than or equal to 0.5.
The  existence of optimal temperature control for preserving a nuclear medium is seen as Area $ \textcircled{\scriptsize 1}$, where the momentum of the two colliding nuclei is also well conserved with several \% accuracy.
The tetra-neutron is calculated to survive only in these cases; tetra neutrons break up into separated neutrons if the energy is higher (Area $\textcircled{\scriptsize 3}$), while tetra neutron is mixed up with the mother nucleus if the energy is lower (Area $\textcircled{\scriptsize 2}$).
Areas $\textcircled{\scriptsize 2}$ and $\textcircled{\scriptsize 5}$ are classified to the fast charge equilibration.
The two nuclei are not easily mixed up in terms of $N/Z$ ratio (charge distribution) in a large Area $\textcircled{\scriptsize 3}$ in which localized charge distribution tends to survive. 
The multi fragmentation into small fragments are calculated in an Area $\textcircled{\scriptsize 4}$, and therefore the charge distribution tends to be unlocalized. 
Comparing $N/Z$=1 cases (thick lines) to $N/Z$=7/5 cases (dashed lines), the neutron-richness and thus the nuclear symmetry energy effects play a role to a certain degree.  
Eventually we conclude that soliton existence and therefore the tetra-neutron independence are mass dependent in the sense of inclusive mother nuclei and its existence can be confirmed only for small mass cases ($Z \le 8$).

\begin{figure}
\includegraphics[width=70mm]{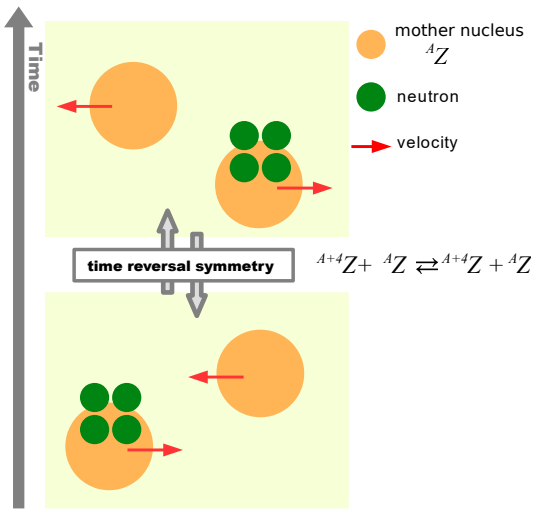} 
\caption{ \label{fig0}
(Color online) A conservation during ion collisions in which the initial and final nuclei are the same in terms of inclusive neutron and proton numbers, and of the amplitudes of relative momentum of nucleus.
Tetra-neutron can be broken depending on the mother nucleus and the momentum.
}
\end{figure}

What is clarified by the energy and mass dependent soliton preservation being labeled by $Z$, $A$ and $E_K$ (corresponding to the Area $\textcircled{\scriptsize 1}$ of Fig.~\ref{fig11}) is associated with the time reversal symmetry, in which there is no nucleon transfer, no fragmentation, no nucleon emission, and no internal excitation.
The Area $\textcircled{\scriptsize 1}$ is associated with the time-reversal symmetry.
The time reversal symmetry arises from the energy conservation (Fig.~\ref{fig0}), according to  Noether's theorem.
It is worth noting that total energy conservation is valid due to the unitary time evolutions described the TDDFT, while the time reversal symmetry is not necessarily true in the collision dynamics described by the TDDFT because of the appearance of internal single particle excitations and the time-odd components of nuclear energy density functional, as well as dissipation by particle emission at higher energies.
The present study suggests the conditional recovery of time-reversal symmetry of certain nuclear medium by a set of light ions ($Z \le 8$, $A \le 2Z+4$) with a certain kinetic energy range. 

The concept of low-energy solitonic core or solitonic layer of stellar object is proposed based on the conditional nuclear conservation.
Some compact stars are mainly composed of $\alpha$-particle ($^{4}$He);
since the core temperature of compact stars can be from 10 MeV to 20 MeV (e.g., core temperature of white dwarf), $\alpha$-particle is less reactive to capture neutrons or to form another fused system (Fig.~\ref{fig4}), and $\alpha$-particles are rather well preserved.
$\alpha$-particles are not reactive for certain temperature range, even if sufficient energy for fusion reactions as much as 10 MeV per nucleon is provided.
Meanwhile the soliton star is a long-standing concept whose existence is proposed in association with the relativistic gravitational force field~\cite{87lee} (for a textbook of bosonic stars, soliton stars and the related topics, see \cite{17liebling}).
The energy of such proposed soliton stars (comparable to the energy of black holes) is much higher than the present low-temperature solitonic core.
Since the temperature range of low-energy soliton propagation is limited from 5 to 15 MeV (Area  $ \textcircled{\scriptsize 1}$ of Fig. 4) in the present cases, ``low-energy'' cold soliton star can exist at only in limited situations, but the solitonic core or solitonic layer of the stellar object is likely to be realized instead. 

\section{Conclusion}

In conclusion solitonic transportation is shown for many-nucleus systems.
A competition between the soliton propagation (a kind of nonlinearity) and the charge equilibration (a kind of many-body property) is essential to the realization of solitons under momentum equilibration.
Such a competition is shown to be fully parametrized by the temperature.
The results are summarized by the following itemization.
With focusing on the energy-dependent existence of solitonic transportation in low-energy heavy-ion collisions, a systematic TDDFT calculation shows the following statements:
\begin{itemize}
\item the conditional recovery of time-reversal symmetry is suggested to be valid in many nucleus systems, if they are composed of light nuclei ($Z \le 8$);
\item the independent motion of the tetra-neutron is shown to exist only if the temperature is between 5 and 15 MeV; 
\item from a technological point of view, existence and non-existence of temperature control for preserving nuclear medium is shown to be dependent mainly on the mass of components without being much influenced by the neutron-richness of components;
\item from an astrophysical point of view, the concept of low-energy solitonic state for the core of compact star is proposed.
\end{itemize}
In particular, a tetra-neutron will find it hard to survive in $^{4}$He+$^{8}$He reaction if the collision energy is less than 20 MeV per nucleon in the center-of-mass frame and therefore 90 MeV per nucleon in the laboratory frame (for the conversion from center-of-mass frame to laboratory frame, see an unlabeled equation between Eqs.~(1) and (2) of \cite{13iwata-02}).
A tetra-neutron search in $^{4}$He+$^{8}$He will not show us only some findings for neutron-rich medium, but also the breaking mechanism of time reversal symmetry in femto-meter scale quantum systems.
Using the nuclear equations of state to be discovered, the control could be generalized to pressure control or volume control. \vspace{6mm}\\

Numerical computation was carried out at Yukawa Institute Computer Facility of Kyoto University, and a workstation system at Tokyo Institute of Technology.
This work was supported by JSPS KAKENHI Grant No. 17K05440, and UK STFC grant no. ST/P005314/1.

\end{document}